\documentclass{ws-ijmpcs}

\begin{document}

\markboth{Guillaume Beuf}
{High-energy factorization and evolution with improved kinematics}

\catchline{}{}{}{}{}
%
%%%%%%%%%%%%%%%%%%%%%%%%%%%%%%%%%%%%%%%%%%%%%%%%%%%%%%%%%%%%%%%%%%%%

\title{HIGH-ENERGY FACTORIZATION AND EVOLUTION WITH IMPROVED KINEMATICS}

\author{GUILLAUME BEUF}

\address{Department of Physics,
Brookhaven National Laboratory,\\
Upton, NY 11973, USA\\
gbeuf@quark.phy.bnl.gov}

\maketitle

%\begin{history}
%\received{Day Month Year}
%\revised{Day Month Year}
%\end{history}

\begin{abstract}
The high-energy factorization and the associated B-JIMWLK or BK evolution equations are presented, using the example of DIS structure functions. The necessity of taking gluon saturation into account is discussed, and also the various approximations underlying high-energy factorization.
The appearance of large NLL corrections in such a framework with or without gluon saturation is recalled, and their physical origin is explained. Finally, old and new results are presented about the resummation of some of those large corrections, related to kinematical approximations.

\keywords{High-energy QCD; gluon saturation; factorization and evolution.}
\end{abstract}

\ccode{PACS numbers: }

\section{Introduction}	

QCD hard processes are described within the improved parton model, thanks to the collinear factorization. In that kinematical regime, only a few quasi on-shell quarks and gluons per hadron are relevant for the dynamics, and thus hadrons are appropriately described by parton distribution functions.

However, this picture breaks down in the high-energy limit or equivalently at low Bjorken-$x$. Indeed, due to the enhancement of soft gluon {\it Bremsstrahlung}, the gluon occupation number reaches non-perturbatively large values at low enough $x$. Gluons in this regime are better described as a semiclassical field rather than as a dilute gas of partons. The growth of the gluon occupation number is then tamed by the non-linearity of the Yang-Mills equations. This is the phenomenon of gluon saturation.\cite{Gribov:1984tu,Mueller:1985wy,McLerran:1993ni,McLerran:1993ka,McLerran:1994vd}

This phenomenon is taken into account in various formalisms which aim at describing dense-dilute and/or dense-dense observables. In the former case, the low-$x$ gluons of a hadron or nucleus target are probed by a dilute projectile. Typical examples of that case are Deep Inelastic Scattering (DIS) observables at low-$x$ and forward physics in hadron-hadron or hadron-nucleus collisions.
By contrast, dense-dense observables involve the collision of the low-$x$ gluon field of two hadrons or nuclei against each other. The inclusive $n$-particle production at mid-rapidity in heavy ion or hadron-hadron collision at high-energy enters in that category, as well as the $n$-point correlators of the energy-momentum released at mid-rapidity in such collisions.

In this contribution, I focus on the paradigmatic example of DIS, whereas dense-dense processes are discussed by Venugopalan in his contribution.\cite{Raju} Note that both for dense-dilute and dense-dense observables, the high-energy leading logs are resummed by the same B-JIMWLK equations\cite{Balitsky:1995ub,JalilianMarian:1997jx,JalilianMarian:1997gr,JalilianMarian:1997dw,Kovner:2000pt,Weigert:2000gi,Iancu:2000hn,Iancu:2001ad,Ferreiro:2001qy} acting on Wilson line operators, first derived without explicit reference to the projectile. Hence, there is a factorization property valid at leading logarithmic accuracy for several very different observables. In the large $N_c$ limit, the B-JIMWLK equations applied to a dipole of Wilson lines operators reduce to the much simpler BK equation.\cite{Balitsky:1995ub,Kovchegov:1999yj,Kovchegov:1999ua} Both the B-JIMWLK equations and the BK equation reduce to the BFKL equation\cite{Lipatov:1976zz,Kuraev:1977fs,Balitsky:1978ic} in the dilute limit for the target.

Recently, Balitsky and Chirilli have calculated\cite{Balitsky:2008zza,Balitsky:2009xg} the evolution equation resumming the next to leading logarithms (NLL) in the particular case of dipoles of Wilson lines operators (and thus the NLL BK equation), as well as the NLO impact factor\cite{Balitsky:2010ze} (or high-energy coefficient-functions) for DIS structure functions. These calculations show that the high-energy factorization remains valid at NLO/NLL accuracy, at least in the case of DIS structure functions. This high-energy factorization generalizes the dipole factorization\cite{Nikolaev:1990ja}, valid at LO only.    Hence, high-energy factorization is now a good candidate for a factorization property valid to all orders, analog to collinear and TMD factorizations.

However, the NLL BK equation is plagued by problems inherited from its linearized analog, the NLL BFKL equation.\cite{Fadin:1998py,Ciafaloni:1998gs} In both equations, the NLO contributions to the kernel are large, especially in the collinear and anti-collinear kinematical regimes, signaling a breakdown of the weak coupling expansion in most of the available phase-space. Hence, in order to obtain stable and reliable solutions of the NLL evolution equations, one needs to perform appropriate resummations, already done in the BFKL case.\cite{Ciafaloni:1999yw,Ciafaloni:2003rd,Altarelli:2005ni}

In the next section, the derivation of the high-energy factorization is outlined in the context of light-front quantization, and the underlying assumptions are discussed in detail. In the section \ref{sec:largeCorr}, the large corrections appearing in NLL BFKL and NLL BK equations and their physical origin are discussed. Progresses towards the resummation of these corrections to the NLL BK equation are also reviewed. In the section \ref{sec:KinConst}, preliminary results about the resummations of the large corrections of kinematical origin are introduced, with emphasis on the factorization scheme dependence of these effects.

\section{High-Energy Factorization for DIS}\label{HE_fact_DIS}	

The high-energy factorization for DIS can be obtained both from covariant or light-front approaches to QCD. The formalism used by Balitsky and Chirilli and discussed in their respective contributions,\cite{Ian,Giovanni} is fully covariant and thus suitable to perform systematically higher order calculations. As an alternative, let us use a light-front formalism, which allows to make the physical content of the calculations more transparent at leading order or at leading logarithmic accuracy. This formalism is essentially the QCD version of the one constructed by Bjorken, Kogut and Soper for QED.\cite{Bjorken:1970ah}

Let us choose a frame such that the projectile moves mainly along the $x^+$ direction and the target mainly along $x^-$, and such that the target is much more boosted than the projectile. In the high-energy (or low $x$) limit, the target is more and more Lorentz contracted along $x^+$, so that the $x^+$ time interval around $x^+=0$ needed for the projectile to pass through the target becomes arbitrarily short. Hence, the forward scattering process related to DIS by the optical theorem happens in three steps
\begin{itemize}
  \item the projectile propagates in the vacuum form $x^+=-\infty$ to $0$,
  \item the projectile interacts with the target at $x^+=0$,
  \item the projectile propagates in the vacuum form $x^+=0$ to $+\infty$,
\end{itemize}
up to contributions suppressed by powers of the Lorentz $\gamma$ factor of the target.

The formalism of light-front wave functions allows to describe the state of the projectile just before or just after the collision. A light-front wave function is indeed the overlap of the physical state of the projectile with a given Fock state at $x^+=0$.
In the DIS case, the projectile is in principle the incoming lepton, but as in covariant calculations, one can factor out the leptonic tensor, and take effectively the projectile to be the exchanged virtual photon, which can be transverse or longitudinal.
At LO, only the quark-antiquark Fock component of the virtual photon projectile contributes to DIS, so that the hadronic contribution to the projectile state writes formally
\begin{equation}
|\textrm{Proj}\rangle^{\textrm{LO}}= \int \textrm{d}PS_{\textrm{q}\bar{\textrm{q}}} \; \; \Psi^{\textrm{LO}}_{\textrm{q}\bar{\textrm{q}}} \; \; |\textrm{q}\bar{\textrm{q}}\rangle\: ,
\end{equation}
with a sum over the phase space and quantum numbers of the quark and antiquark. At NLO, the quark-antiquark-gluon Fock component contributes as well, so that
\begin{equation}
|\textrm{Proj}\rangle^{\textrm{NLO}}= \int \textrm{d}PS_{\textrm{q}\bar{\textrm{q}}} \; \; \Psi^{\textrm{NLO}}_{\textrm{q}\bar{\textrm{q}}} \; \; |\textrm{q}\bar{\textrm{q}}\rangle  +
\int \textrm{d}PS_{\textrm{q}\bar{\textrm{q}}\textrm{g}} \; \; \Psi^{\textrm{NLO}}_{\textrm{q}\bar{\textrm{q}}\textrm{g}} \; \; |\textrm{q}\bar{\textrm{q}}\textrm{g}\rangle \: .
\end{equation}

%, and is divergent in the limit of vanishing light-cone momentum $k^+$ of the gluon.

In the high-energy limit, the various partons present in a Fock state of the projectile scatter independently on the target, up to power suppressed corrections. Indeed, the time interval for the interaction between the target and the projectile goes to zero when boosting more and more the target, and also the typical time scale for interactions within the projectile is Lorentz dilated when boosting the projectile.

Following the Color Glass Condensate (CGC) formalism\cite{McLerran:1993ni,McLerran:1993ka,McLerran:1994vd,Iancu:2000hn,Iancu:2001ad,Ferreiro:2001qy}, let us model the target by a random classical gluon field, obeying some classical statistical distribution. I will discuss the validity of this assumption later on. Then, we need to know how a parton scatters in the high-energy limit on a classical gluon field shrinking to a shockwave at $x^+=0$. When passing through such a shockwave, the transverse position of a parton has no time to change. Hence, it is convenient to write the Fock state basis in a mixed representation where the transverse position rather than the transverse momentum of the partons is specified, in addition to their light-cone momentum $k^+$. Both the probability of helicity flip of the parton and the probability of loss of a non-negligible fraction of his light-cone momentum $k^+$, by scattering on the gluon shockwave, are power suppressed in the high-energy limit. Hence, the scattering of independent partons on the classical gluon shockwave reduces to the so-called eikonal scattering, which is diagonal in the Fock state basis except with respect to color indices. A parton is thus only color rotated when passing through the shockwave field. This color rotation is given by a Wilson line in the appropriate representation, defined along the trajectory of the parton through the target gluon field, with fixed transverse position $\mathbf{x}$, noted for example $U(\mathbf{x})$ in the case of a quark.

In particular, the S-matrix for the forward scattering of a free quark-antiquark color singlet dipole (at transverse positions $\mathbf{x}_0$ and $\mathbf{x}_1$) on a classical gluon shockwave is the identity matrix in Fock space times the factor
\begin{equation}
S_{\mathbf{x}_0 \mathbf{x}_1}= \frac{1}{N_c} \textrm{tr}\left(U(\mathbf{x}_0) U^\dag(\mathbf{x}_1)\right)\: .
\end{equation}
Using Fierz identities, the analog S-matrix factor for a quark-antiquark-gluon color singlet can be written as
\begin{equation}
S_{\mathbf{x}_0 \mathbf{x}_1 \mathbf{x}_2}= \frac{1}{N_c^2\!-\!1}\left[ \textrm{tr}\left(U(\mathbf{x}_0) U^\dag(\mathbf{x}_2)\right) \; \textrm{tr}\left(U(\mathbf{x}_2) U^\dag(\mathbf{x}_1)\right)-\frac{1}{N_c} \textrm{tr}\left(U(\mathbf{x}_0) U^\dag(\mathbf{x}_1)\right)\right]  \: ,
\end{equation}
where $\mathbf{x}_2$ is the transverse position of the gluon, so that
\begin{equation}
S_{\mathbf{x}_0 \mathbf{x}_1 \mathbf{x}_2}= \frac{N_c^2}{N_c^2\!-\!1}\left[S_{\mathbf{x}_0 \mathbf{x}_2}\, S_{\mathbf{x}_2 \mathbf{x}_1}-  \frac{1}{N_c^2} S_{\mathbf{x}_0 \mathbf{x}_1}\right]\: .
\end{equation}

All in all, the projectile-target total cross section writes at LO as
\begin{equation}
\sigma^{\gamma^*}_{\textrm{tot}} = 2 \int \textrm{d}PS_{\textrm{q}\bar{\textrm{q}}} \; \; \left|\Psi^{\textrm{LO}}_{\textrm{q}\bar{\textrm{q}}}\right|^2 \; \; \Big[1-\big< S_{\mathbf{x}_0 \mathbf{x}_1}\big> \Big]        \: ,\label{LO_cross_section}
\end{equation}
with a different wave function $\Psi^{\textrm{LO}}_{\textrm{q}\bar{\textrm{q}}}$ whether the virtual photon projectile is transverse or longitudinal, and where $\big< \dots \big>$ represents the statistical average over the target gluon field.
And at NLO, one expects the total cross section to be
\begin{equation}
\sigma^{\gamma^*}_{\textrm{tot}} = 2 \int \textrm{d}PS_{\textrm{q}\bar{\textrm{q}}} \; \; \left|\Psi^{\textrm{NLO}}_{\textrm{q}\bar{\textrm{q}}}\right|^2 \; \; \Big[1-\big< S_{\mathbf{x}_0 \mathbf{x}_1}\big> \Big]
+
2 \int \textrm{d}PS_{\textrm{q}\bar{\textrm{q}}\textrm{g}} \; \; \left|\Psi^{\textrm{NLO}}_{\textrm{q}\bar{\textrm{q}}\textrm{g}}\right|^2 \; \; \Big[1-\big<S_{\mathbf{x}_0 \mathbf{x}_1 \mathbf{x}_2}\big> \Big]
 \: ,
\end{equation}
and so on at higher orders. Using the probability conservation relation
\begin{equation}
\left|\Psi^{\textrm{NLO}}_{\textrm{q}\bar{\textrm{q}}}\right|^2+\int \textrm{d}PS_{\textrm{g}} \; \; \left|\Psi^{\textrm{NLO}}_{\textrm{q}\bar{\textrm{q}}\textrm{g}}\right|^2 = \left|\Psi^{\textrm{LO}}_{\textrm{q}\bar{\textrm{q}}}\right|^2 \: ,
\end{equation}
the total cross section at NLO can be rewritten as
\begin{eqnarray}
\sigma^{\gamma^*}_{\textrm{tot}} &=& 2 \int \textrm{d}PS_{\textrm{q}\bar{\textrm{q}}} \; \; \left|\Psi^{\textrm{LO}}_{\textrm{q}\bar{\textrm{q}}}\right|^2 \; \; \Big[1-\big< S_{\mathbf{x}_0 \mathbf{x}_1}\big> \Big]\nonumber\\
& &+\frac{2 N_c^2}{N_c^2\!-\!1} \int \textrm{d}PS_{\textrm{q}\bar{\textrm{q}}\textrm{g}} \; \; \left|\Psi^{\textrm{NLO}}_{\textrm{q}\bar{\textrm{q}}\textrm{g}}\right|^2 \; \; \Big[\big< S_{\mathbf{x}_0 \mathbf{x}_1}\big>-\big<S_{\mathbf{x}_0 \mathbf{x}_2}\, S_{\mathbf{x}_2 \mathbf{x}_1}\big> \Big]\label{NLO_cross_section}
 \: ,
\end{eqnarray}
separating the LO and NLO contributions from each other.

However, this is not the end of the story. Indeed, the operators constructed from light-like Wilson lines such as $S_{\mathbf{x}_0 \mathbf{x}_1}$ suffer from rapidity divergences, coming from gluons of large $k^+$ and/or vanishing $k^-$ belonging to the field of target. And the usual logarithmic divergence of soft gluon {\it Bremsstrahlung} shows up in the NLO contribution as
\begin{equation}
\textrm{d}PS_{\textrm{q}\bar{\textrm{q}}\textrm{g}} \; \left|\Psi^{\textrm{NLO}}_{\textrm{q}\bar{\textrm{q}}\textrm{g}}\right|^2 \propto  \frac{\textrm{d} k^+}{k^+} \quad \textrm{for small $k^+$ of the gluon.}
\label{brems_log_div_kplus}
\end{equation}
These two divergences are both regulated when one avoids double counting of gluons, thanks to a longitudinal cut-off. For example, let us take a cut-off $k^+_c$ in $k^+$, and consider gluons with $k^+> k^+_c$ only as part of the projectile, in order to regulate the integration with the weight \eqref{brems_log_div_kplus}, and gluons with $k^+< k^+_c$ only as part of the target semi-classical field, so that $\big< S_{\mathbf{x}_0 \mathbf{x}_1}\big>$ and $\big<S_{\mathbf{x}_0 \mathbf{x}_2}\, S_{\mathbf{x}_2 \mathbf{x}_1}\big>$ become $k^+_c$-dependent.

The evolution with $k^+_c$ of $\big< S_{\mathbf{x}_0 \mathbf{x}_1}\big>_{k^+_c}$, $\big<S_{\mathbf{x}_0 \mathbf{x}_2}\, S_{\mathbf{x}_2 \mathbf{x}_1}\big>_{k^+_c}$ and similar expectation values is obtained by integrating out QCD quantum corrections slice by slice in $k^+$. At leading log accuracy, one gets
\begin{equation}
k^+_c \partial_{k^+_c} \big< S_{\mathbf{x}_0 \mathbf{x}_1}\big>_{k^+_c}= \frac{N_c \alpha_s}{2 \pi^2} \int \textrm{d}^2 \mathbf{x}_2\: \frac{x_{01}^2}{x_{02}^2\, x_{21}^2}\: \Big[\big<S_{\mathbf{x}_0 \mathbf{x}_2}\, S_{\mathbf{x}_2 \mathbf{x}_1}\big>_{k^+_c}-\big< S_{\mathbf{x}_0 \mathbf{x}_1}\big>_{k^+_c}\Big]\label{eqBal1}
\end{equation}
where $x_{ij}^2=(\mathbf{x}_i\!-\!\mathbf{x}_j)^2$, and similar other coupled equations forming Balitsky's infinite hierarchy,\cite{Balitsky:1995ub} involving expectations values of operators with an arbitrary number of Wilson lines. The JIMWLK equation\cite{JalilianMarian:1997jx,JalilianMarian:1997gr,JalilianMarian:1997dw,Kovner:2000pt,Weigert:2000gi,Iancu:2000hn,Iancu:2001ad,Ferreiro:2001qy} is equivalent to this full hierarchy, and gives the evolution with $k^+_c$ of the target gluon field probability distribution $\big< \dots\big>_{k^+_c}$. The fact that the quantum corrections to the statistical distribution of gluons fields can be absorbed by evolution of the statistical distribution is a crucial self-consistency check for the CGC formalism.

Alternatively, the same equation \eqref{eqBal1} can be derived\cite{Kovchegov:1999yj} by extracting the soft-gluon logarithmic divergence coming from the photon light-cone wave function in \eqref{NLO_cross_section}, and requiring the DIS cross section to be independent of the cut-off $k^+_c$, playing the role of a longitudinal factorization scale. The consistency of the two approaches shows the validity of the high-energy factorization formulae \eqref{LO_cross_section} and \eqref{NLO_cross_section} for DIS once the cut-off $k^+_c$ is applied, since the B-JIMWLK evolution properly resums the high-energy leading logs.

In the large $N_c$ limit, one has $\big<S_{\mathbf{x}_0 \mathbf{x}_2}\, S_{\mathbf{x}_2 \mathbf{x}_1}\big>_{k^+_c} \rightarrow \big<S_{\mathbf{x}_0 \mathbf{x}_2}\big>_{k^+_c}\, \big<S_{\mathbf{x}_2 \mathbf{x}_1}\big>_{k^+_c}$, so that \eqref{eqBal1} becomes a closed equation, called Balitsky-Kovchegov (BK) equation.\cite{Balitsky:1995ub,Kovchegov:1999yj,Kovchegov:1999ua}

When using the NLO factorization formula \eqref{NLO_cross_section}, by consistency, one should probably resum not only the high-energy leading logs, but also the next-to-leading ones. This is performed by the NLL extension of the equation \eqref{eqBal1}, derived recently.\cite{Balitsky:2008zza,Balitsky:2009xg}

Let us come back to the main assumption of the CGC formalism, {\it i.e.} that the target can be modeled by a random classical gluon field. First, we have seen the self-consistency of this assumption thanks to the JIMWLK equation. Second, when the gluon occupation number is large, it is large compared to commutators, so that the physics becomes effectively classical and the CGC assumption is justified. Due to Lorentz contraction, the valence partons of a relativistic large nucleus are densely packed, so that the CGC formalism applies to large nuclei as soon as relativistic effects are important.
Moreover, the B-JIMWLK evolution leads to a more and more packed target, so that all this formalism is always valid at high enough energy, whatever is the target, even if in intermediate stages of the high-energy evolution, the target is dilute. Finally, the dilute limit (or two-gluon approximation) of the equation \eqref{eqBal1} and of its NLL extension correctly reproduces the LL and NLL BFKL equations respectively, so that at this accuracy, no information about the dilute regime is lost when taking the semi-classical target approximation. However, this approximation for the target field might prevent to reproduce the full NNLL BFKL equation in the dilute limit.

\section{Large NLL corrections and collinear resummations}\label{sec:largeCorr}

The BFKL equation receives large corrections at NLL,\cite{Fadin:1998py,Ciafaloni:1998gs} giving a pathological behavior to its solutions and signaling a breakdown of the corresponding formalism. Soon after this discovery, it has been realized that these issues come entirely from the collinear and anti-collinear kinematical regimes and not from the Regge kinematical regime in which the BFKL equation is derived. The solution is then to perform resummations\cite{Ciafaloni:1999yw,Ciafaloni:2003rd,Altarelli:2005ni} in the collinear and anti-collinear regimes, in order to extend the validity of the NLL BFKL toward these regimes, which are in principle govern by DGLAP physics. Several successful resummations schemes have been proposed, which differ only by inessential higher order contributions. The general idea behind these schemes is to add appropriate NNLL and higher order terms to the NLL BFKL equation in such a way that the (N)LO DGLAP equation is reproduced in the collinear and anti-collinear limits.

The physical origin of the large corrections has also been understood. They can be classified as follows.
\begin{itemize}
\item{Running coupling scale:} The NLL BFKL kernel contains terms which are the beginning of the expansion of the running coupling. When the coupling is taken to run, but not with the correct scale, some of these terms stay in the NLL kernel instead of being resummed into the running coupling. If the running coupling scale differs in the collinear or anti-collinear regime from the one dictated by DGLAP physics, the left-over terms in the kernel are particularly large.
\item{Non-singular part of DGLAP splitting function:} The LL BFKL equation contains the singular part of the LO DGLAP gluon splitting function both in the collinear and anti-collinear regimes, but not the non-singular part. The non-singular part shows up in an expanded way, term after term when going to NLL BFKL and further. This expanded form appears to be pathological, and one should instead include the full non-singular part together with the singular part of the splitting function in a modified version of the LL BFKL equation.
\item{Kinematical constraint:} The last large corrections, which are parametrically the largest ones, come from a bad treatment of kinematics when taking the high-energy limit too strictly. They can be resummed by imposing some kinematical constraint\cite{Ciafaloni:1987ur,Andersson:1995ju,Kwiecinski:1996td} in the LL BFKL equation, or by shifting collinear and/or anticollinear poles of the kernel in Mellin space.\cite{Salam:1998tj}
\end{itemize}

The corrections of the third type depend on the factorization scheme, in particular on the precise definition of the evolution variable for the BFKL equation. They are however identical in any Yang-Mills theory.
By contrast, the corrections of the first two types are high-energy factorization scheme independent. But they vary from theory to theory. In particular, they are both absent in ${\cal N}=4$ SYM theory.

Since the linearization of the NLL BK equation gives the NLL BFKL equation, those large corrections are also present in the NLL BK equation. The nonlinearity related to gluon saturation in the BK equation cannot help stabilize the solutions in the presence of those large corrections, as shown by toy-model numerical simulations.\cite{Avsar:2011ds} Hence, it is necessary to perform collinear resummations also for the NLL BK equation before using it for phenomenology. However, the resummation schemes developed for NLL BFKL cannot be used directly for the NLL BK equation, because of the nonlinearity of the latter and of the switch from transverse momentum to transverse position space.

The terms related to running coupling in the NLL BK equation have been analyzed in detail, and an appropriate running coupling prescription is available.\cite{Balitsky:2006wa} Concerning the non-singular part of the gluon splitting function, only a very rough treatment is available,\cite{Gotsman:2004xb} so that more work on this issue is desirable. Let us now discuss the kinematical issues in the BK context.

\section{Resummation via kinematical improvement}\label{sec:KinConst}

The need for the kinematical constraint in the evolution equation was first discovered in a covariant formalism, in a quite technical calculation aiming at the description of other effects like angular ordering.\cite{Ciafaloni:1987ur}
However, the origin of the kinematical constraint and of the related large NLL corrections can be understood very clearly in light-front perturbation theory, as shown in Ref.~\refcite{Motyka:2009gi}.

The easiest derivation of the LL BFKL and BK equations is performed by calculating the leading logarithmic contributions to the squared wave functions $\left|\Psi^{\textrm{N}^n\textrm{LO}}_{\textrm{q}\bar{\textrm{q}}\textrm{g}\cdots \textrm{g}}\right|^2$ for the quark antiquark and $n$ gluons Fock component of the projectile in the multi-Regge kinematics.\cite{Mueller:1993rr,Kovchegov:1999yj} In light-front perturbation theory, the kinematical issues are only due to the usual implementation of the multi-Regge kinematics on the energy denominators, coming with each additional gluon. In this context, the multi-Regge kinematics is the assumption that the $k^+$ of the partons are strongly ordered\footnote{Here, the gluons are ordered along the gluon cascade from the closest to the projectile, $2$, to the closest to the target, $n+1$. $k^+_1$ is the momentum of the antiquark, and $q^+$ the one of the virtual photon.}
\begin{equation}
q^+ > k^+_1,\: q^+\!-\!k^+_1 \gg k^+_2 \gg  \cdots \gg  k^+_{n+1} \; ,\label{kplus_ordering}
\end{equation}
and that simultaneously their transverse momentum are of the same order
\begin{equation}
 Q^2 \simeq \mathbf{k}^2_1 \simeq \mathbf{k}^2_2 \simeq  \cdots \simeq  \mathbf{k}^2_{n+1} \; .\label{kt_equal}
\end{equation}
If one assumes only the $k^+$ ordering \eqref{kplus_ordering}, then for each diagram, the energy denominator associated with the emission of the gluon $j$, with $2\leq j\leq n+1$, reduces to
\begin{equation}
\frac{Q^2}{2 q^+}+ \frac{q^+\, \mathbf{k}^2_1}{2 k^+_1 (q^+\!-\!k^+_1)}+ \frac{\mathbf{k}^2_2}{2 k^+_2}+ \cdots +  \frac{\mathbf{k}^2_j}{2 k^+_j}\, .\label{Energy_denom}
\end{equation}
Then, in the strict multi-Regge kinematics \eqref{kplus_ordering}-\eqref{kt_equal}, one can approximate each energy denominator by its last term
\begin{equation}
\frac{Q^2}{2 q^+}+ \frac{q^+\, \mathbf{k}^2_1}{2 k^+_1 (q^+\!-\!k^+_1)}+ \frac{\mathbf{k}^2_2}{2 k^+_2}+ \cdots +  \frac{\mathbf{k}^2_j}{2 k^+_j}\simeq \frac{\mathbf{k}^2_j}{2 k^+_j}=k^-_j\, .\label{Energy_denom_approx}
\end{equation}
That approximation is crucial to the derivation of the LL BFKL and BK equations. When the high-energy factorization is formulated with a cut-off in $k^+$, as presented in section \ref{HE_fact_DIS}, the $k^+$-ordering \eqref{kplus_ordering} of the gluon cascades generated by the BFKL and BK equations is automatic. However, the integration over transverse momentum or position in the kernel of the LL BFKL and BK equations is not restricted, so that it contains contributions badly violating \eqref{kt_equal} and thus outside the multi-Regge kinematics. More precisely, once the ordering \eqref{kplus_ordering} is satisfied, the approximation \eqref{Energy_denom_approx} is good typically for gluon cascades in the multi-Regge kinematics \eqref{kt_equal} or in the anti-collinear kinematics, {\it i.e.} with $\mathbf{k}^2_j$ growing with $j$, but not in the deep collinear regime, {\it i.e.} with $\mathbf{k}^2_j$ decreasing fast enough with $j$.

From this discussion, ones predicts that the $k^+$-ordered BFKL and BK equations are less reliable in the deep collinear regime, which is especially relevant for DIS, than in the rest of the available phase-space. This observation explains why the Mellin transform of the NLL kernel of the $k^+$-ordered BK equation\cite{Balitsky:2008zza} contains a large and unphysical collinear triple pole, but no anti-collinear triple pole.

All this discussion is however factorization scheme dependent. There are indeed other ways to regularize the rapidity divergence of Wilson line operators and the soft divergence \eqref{brems_log_div_kplus}. For example, one can replace the cut-off in $k^+$ by a cut-off in $k^-$. In that case, the BFKL and BK equations implement automatically the $k^-$-ordering
\begin{equation}
\frac{Q^2}{2 q^+}+ \frac{q^+\, \mathbf{k}^2_1}{2 k^+_1 (q^+\!-\!k^+_1)} \ll k^-_2 \ll  \cdots \ll  k^-_{n+1} \; ,\label{kminus_ordering}
\end{equation}
so that approximations of the type \eqref{Energy_denom_approx} are good. However, this time, the approximation of the energy denominators by the expression \eqref{Energy_denom} is not always valid. In fact, we are in the inverse situation: the approximation of energy denominators by the $k^-$ of the last gluon is in general valid except in the deep anti-collinear regime. Hence, a spurious anti-collinear triple pole is expected in the Mellin transform of the $k^-$-ordered NLL BFKL and BK kernels, but no collinear triple pole.

In fact, whatever factorization scheme is chosen, the approximation of the energy denominators by the $k^-$ of the last gluon is valid if and only if both the $k^+$ and $k^-$-ordering \eqref{kplus_ordering}-\eqref{kminus_ordering} are satisfied, whereas the assumption \eqref{kt_equal} is unnecessary. Moreover, one can show that if either \eqref{kplus_ordering} or \eqref{kminus_ordering} is not satisfied, no leading logarithmic contribution is generated.\cite{me} Hence, the BFKL and BK equations should actually resum precisely the gluon cascades ordered both in $k^+$ and $k^-$,\cite{Motyka:2009gi,me} whereas the usual assumption \eqref{kt_equal} is unnecessary and misleading. In any factorization scheme, imposing both $k^+$ and $k^-$ ordering allows to resum into a modified LL equation the largest unphysical collinear or anti-collinear contributions appearing in the kernel at higher logarithmic orders, corresponding in Mellin space to triple poles at NLL, quintuple poles at NNLL and so on.

Only the $k^+$-ordering or the $k^-$-ordering, not both, can be made automatic by a choice of high-energy factorization scheme. The other one has to be included as a modification of the LL kernel, by restricting the integration over transverse momentum or position to the domain allowing to satisfy both \eqref{kplus_ordering} and \eqref{kminus_ordering}. For the BFKL equation in momentum space, this restriction of the integration domain is precisely the kinematical constraint found in Ref.~\refcite{Ciafaloni:1987ur}.

So far, we have shown the necessity of the kinematical constraint in the BFKL and BK evolutions by an analysis in transverse momentum space. However, evolutions equations including gluon saturation like the BK and JIMWLK equations are more easily written in transverse position space. Hence, one has to translate the kinematical constraint to a restriction in position space. A first implementation of the kinematical constraint in position space for the BFKL and BK equations has been proposed in Ref.~\refcite{Motyka:2009gi}. I have recently clarified\cite{me} some issues related to this first implementation, in particular by doing a more careful treatment of virtual corrections and by analyzing systematically the factorization scheme dependence of the kinematical constraint. In the end, one gets a modified BK evolution equation with memory effects and a restricted phase-space for gluon emission which gets progressively wider in the course of the evolution.\cite{me}

\section*{Acknowledgments}

This manuscript has been authored under the Contract No. \#DE-AC02-98CH10886 with the
U.S. Department of Energy.

%\appendix

%\begin{thebibliography}{000} %for 3 digits
%\begin{thebibliography}{00}  %for 2 digits

\end{document}